\documentstyle[11pt,aasms4]{article}

\received{}
\accepted{}
\journalid{}{}
\articleid{}{}

\slugcomment{Accepted for ApJ Letters}

\lefthead{Martin et al.}
\righthead{The First L-type Brown Dwarf in the Pleiades}

\begin{document}

\title{The First L-type Brown Dwarf in the Pleiades}

\author{E.L. Mart\'\i n and G. Basri}
\affil{Astronomy Department, University of California,
    Berkeley, CA 94720}

\and

\author{M.R. Zapatero-Osorio, R. Rebolo and R.J. Garc\'\i a L\'opez}
\affil{Instituto de Astrof\'\i sica de Canarias, 38200 La Laguna, Spain}

\centerline{e-mail addresses: ege@popsicle.berkeley.edu, basri@soleil.berkeley.edu} 
\centerline{mosorio@ll.iac.es,rrl@ll.iac.es,rgl@ll.iac.es}

\begin{abstract}
We have obtained low-resolution optical spectra of three faint  
brown dwarf candidates in the Pleiades open cluster. 
The objects observed are Roque~12 ($I_C$=18.5), 
Roque~5 ($I_C$=19.7) and Roque~25 ($I_C$=21.2). 
The spectrum of Roque~25 does not show the strong TiO bandheads that characterize the optical 
spectra of M-type stars, but molecular bands of CaH, CrH and VO are 
clearly present. We classify Roque~25 as an early L-type  
brown dwarf.  Using current 
theoretical evolutionary tracks   
we estimate that the transition from  
M-type to L-type in the Pleiades (age$\sim$120~Myr) 
takes place at T$_{\rm eff}\sim$2200~K or M$\sim$0.04M$_\odot$.  
Roque~25 is a benchmark brown dwarf in the Pleiades because 
it is the first known 
one that belongs to the L-type class. It provides evidence  
that the IMF extends down to about 0.035M$_\odot$, and serves 
as a guide for future deep searches for even less massive young brown 
dwarfs. 

\end{abstract}


\section{Introduction}

Until recently there were no convincing observations of 
brown dwarfs (BDs), but this situation has dramatically changed 
in only a few years. Currently, observations of several  
dozen BDs have been reported. These formerly elusive objects have  
been unambigously identified in the Pleiades  
(Rebolo, Zapatero Osorio \& Mart\'\i n 1995; Basri, Marcy \& Graham \cite{basri96}; Rebolo et al. 1996; 
Zapatero-Osorio et al. 1997; Stauffer, Schultz \& Kirkpatrick 1998),  
$\rho$Oph (Luhman, Liebert \& Rieke 1997), 
as well as in the field (Ruiz, Leggett \& Allard 1997; 
Delfosse et al. 1997; Tinney 1998; Liebert et al. 1998) and 
as companions to stars (Nakajima et al. 1995; Rebolo et al. 1998).  
Optical spectra of the coolest  BD candidates have shown 
that they have properties  
different to those that characterize M-type stars, in particular they 
do not show prominent TiO bands 
(Kirkpatrick et al. 1998; Tinney et al. 1998). 
Thus, it has been proposed to 
use the term ``L-type'' for stars and BDs that 
are too cool to be classified as M-type (Mart\'\i n et al. 1997). 

Searches for BDs in the Pleiades open cluster have been particularly 
fruitful because this cluster offers an ideal combination of 
young age, nearby distance, high metallicity, low reddening, 
richness (over 800 proper motion members) and compactness. 
Three recent deep CCD surveys have provided over 50 photometric Pleiades 
BD candidates. The ITP (International Time Project) CCD survey 
has covered 1~deg$^2$ in the cluster core using $IZ$ filters and 
reaching a completeness limit of $I_C\sim$21.2 (Zapatero Osorio et al. 
1997, 1998). The other two new surveys have been presented by 
Bouvier et al. (1998) and Festin (1998).  The luminosity and mass functions 
inferred from these deep searches suggest that the number of BDs 
in the Pleiades could be of order of a few hundred 
(Zapatero Osorio et al. 1997; 
Mart\'\i n, Zapatero Osorio \& Rebolo 1998; Bouvier et al. 1998). 
The number density of BDs could be similar to that of stars, but they 
probably do not contribute a significant amount of mass to the cluster. 
One of the most important open questions about the IMF is whether there 
is a minimum mass for fragmentation of molecular clouds. 
It is important to characterize the properties of the 
faintest BD candidates discovered by the deep surveys in the Pleiades 
in order to guide the search for even lower mass objects. 
L-type objects in young open clusters are especially interesting 
because they represent very low-mass brown dwarfs. 

The most secure Pleiades BDs are those with kinematic information that 
supports cluster membership and lithium detections that confirm their 
substellar status. The faintest of these BDs are Teide~1 and 
Calar~3 ($I_C$=18.8), 
which have spectral types of M8 (Mart\'\i n, Rebolo \& Zapatero Osorio 
1996) and masses of 0.055$\pm$0.015~M$_\odot$ (Rebolo et al. 1996). 
Fainter Pleiades BD candidates have been reported by the different 
surveys cited above. 
The coolest and faintest BD candidates with published spectra are 
PIZ~1 ($I_C$=19.64; $\sim$M9; 0.048$\pm$0.015~M$_\odot$; 
Cossburn et al. 1997) and Roque~4 
($I_C$=19.75; M9; 0.045$\pm$0.015~M$_\odot$; 
Zapatero Osorio et al. 1997). 
We expect that fainter Pleiades BD candidates 
should be cooler, and thus they should extend the substellar 
spectral sequence into the new L spectral class.  In this Letter we report 
low-resolution spectroscopic observations of three BD candidates 
identified by the ITP team (Zapatero Osorio et al. 1998). 
Two of them have very late M spectral types, and      
the faintest of the three is indeed  
the first L-type object identified in the Pleiades cluster. 

\section{Spectroscopy}

The ``Roque'' BD candidates were discovered in the ITP survey of the 
Pleiades carried out using the 2.5~m Isaac Newton telescope 
(Zapatero Osorio et al. 1997, 1998). In Table~1, we list the names 
and photometry of our program objects.  
Coordinates, additional photometric data and finding 
charts for these and other new Roque objects 
are provided in Zapatero Osorio et al. (1998). 

We observed Roque~25 on 
November 3, 1997, using the LRIS spectrograph attached to the Cassegrain 
focus of the Keck~II telescope.  
The 600/7500 grating with a slit 
width of 1.5 arcseconds and a binning of 2x2 pixels provided a 
resolution of FWHM$\sim$6.1~\AA , and a spectral range 
634.3--891.2~nm. The total integration time was 6600~s. 
We also obtained one 1600~s exposure of Roque~12 in the same night. 
The images were bias subtracted, flat fielded, sky subtracted, 
and variance extracted using routines 
within IRAF\footnote {IRAF is distributed
by National Optical Astronomy Observatory, which is operated by the Association
of Universities for Research in Astronomy, Inc., under contract with the
National Science Foundation.}. We calibrated in wavelength using the 
emission spectrum of NeAr lamps. The 1~$\sigma$ dispersion of the 3rd order 
polynomial fit was 0.12~\AA .  The data  
of Roque~25 were somewhat affected by passing clouds. The telluric water 
bands are stronger than in the spectra of other targets. No flux 
standard was observed in this run 
because we had to close the telescope due to bad weather. 

Additional observations of Roque~25 were obtained 
 in service time on December 17, 1997, with LRIS 
at the Keck~II telescope. The instrumental configuration was slightly 
different than in the November observations. The 600/5000 grating and 
a 1.0 arcsecond slit were used. The spectral range was 639.2--896.5~nm 
and the resolution was FWHM$\sim$5.4~\AA . 
Two exposures of 1800~s were obtained under photometric conditions. 
The flux standard star HZ4 was observed inmediately after Roque~25. 
Using IRAF routines we reduced the spectra as described above, and 
we corrected for instrumental response. 
The final spectra of Roque~12 and Roque~25 are shown in  Figure~\ref{fig1}. 
In the same figure, we also show a low-resolution spectrum of Roque~5, 
that we obtained  at the 4.2~m William Herschel telescope (WHT) 
on December 27, 1997. The red arm of the ISIS spectrograph was used, 
with the R158 grating, a slit width of 2.5 arcsec and a TEK 1024 pix$^2$ 
detector. The resolution was FWHM$\sim$16~\AA , 
and the wavelength range 629.3--925.6~nm. The integration time was 3000~s.  
It was corrected for instrumental response using the flux standard 
G161-B2B observed in the same run. 
In Figure~1, we have also plotted the spectrum of BRI~0021-0214 
obtained by Mart\'\i n et al. (1996), and a spectrum of Kelu~1 that 
we obtained at the WHT on June 16, 1997 using ISIS with the 
R158 grating, a slit width of 1 arcsec and a TEK 1024 pix$^2$ 
detector. The resolution was FWHM$\sim$6~\AA , 
and the wavelength range 633.1--929.0~nm. The integration time was 2400~s. 
We used BD+26 2606 as flux standard.

The spectral types of Roque~12 and Roque~5 were measured using the 
pseudocontinuum indeces defined by Mart\'\i n et al (1996). They should 
be accurate to $\pm$0.5 spectral subclasses. H$\alpha$ emission was detected 
in Roque~12 and Kelu~1. We give the pseudo equivalent widths 
(PEW, measured with respect to the local pseudocontinuum) in Table~1. For 
the other objects we provide upper limits. Li\,{\sc i} 670.8~nm was 
detected in Kelu~1, confirming the detection of 
Ruiz et al. (1997). 
The spectra of the Pleiades objects were  
quite noisy in the lithium region. Consequently, the upper limits to the PEW  
of Li\,{\sc i} are very high. 
The PEW of Li\,{\sc i} among Pleiades BDs are observed to range 
between 0.5 and 2.5~\AA  (Basri et al. 1996; Rebolo et al. 1996; 
Mart\'\i n et al. 1998; Stauffer et al. 1998). 
Our Li\,{\sc i} PEW upper limits are higher than the expected PEW for 
Pleiades BDs because of the low-resolution and poor S/N ratio of 
our data 
in the lithium region. Hence, our non detections do not imply that these 
objects have 
depleted any lithium. The detection of Li\,{\sc i} in Roque~12, 
Roque~5 and Roque~25 should be 
pursued because it would be a final confirmation of their substellar status. 

\section{Discussion}

The position of our program objects   
 in a color-magnitude diagram is presented in Figure~\ref{fig2}. 
Note that the theoretical isochrone for an age of 120~Myr provided 
by Baraffe et al. (1998) provides a fairly good match to the observed cluster 
sequence. 
It is expected that for temperatures cooler than about 2500~K (I$_C$-K$>$3.5) 
the effects of dust become important (Tsuji, Ohnaka \& Aoki 
1996; Jones \& Tsuji 
1997). The Baraffe et al. models use theoretical atmospheres that do not 
include dust opacities (Allard et al. 1997). Thus, the agreement 
between models and observations for the coolest Pleiades BDs should 
improve when dust is incorporated into the models. 
The Pleiades BD sequence is slightly more luminous than the field 
dwarf sequence as expected from the young age of the cluster 
(120~Myr) as determined from the lithium substellar boundary 
(Basri et al. 1996; Mart\'\i n et al. 1998; Stauffer et al. 1998). 
In Figure~2, the sequence 
of field M-dwarfs shifted to the Pleiades distance (127~pc) and 
mean reddening (A$_{\rm V}$=0.12) is represented with a solid line. 
The field dwarfs BRI~0021-0214 and GD~165B are plotted as open squares. 
These objects could be very low-mass (VLM) stars or BDs depending 
on their age, which is still unknown. 
Kelu~1 and the DENIS/2MASS objects do not have known parallaxes yet, and  
cannot be compared with the Pleiades sequence as it can be done 
with the previous two field dwarfs (Tinney, Reid \& Mould 1995). 
The location of Roque~12, Roque~5 and Roque~25 
in Figure~\ref{fig2} indicate substellar 
masses of about 0.060, 0.045 and 0.035~M$_\odot$. 
The effects of dust may lead to slightly lower masses according to 
preliminary computations communicated by I. Baraffe. 
We should point out that our mass estimates depend on the choice 
of evolutionary tracks and the conversion between observables and 
theoretical parameters. We have used the color-temperature scale 
that is produced by the theoretical models. In order to test the 
dependability of our results, we have used other theoretical tracks 
and empirical color-temperature scales with the constraint 
that they should provide a good fit to the Pleiades sequence. 
We found that the tracks of Burrows et al. (1997) 
and D'Antona \& Mazzitelli (1997) 
lead to lower masses for Roque~5 and 25. Thus, our mass estimates 
based on the Baraffe et al. (1998) models are probably conservative.   

We have estimated the 
probability that Roque~25 could be a 
contaminating field object (and not a Pleiades member) using the 
number density of early-L objects (like Kelu~1 and GD~165B) found 
in the DENIS mini-survey (Delfosse et al. 1997).  
The number density of these objects is 0.01--0.005 pc$^{-3}$  
 in the solar vicinity. 
The number density of M8--M9 dwarfs is about 4 times lower 
($\sim$0.0024 pc$^{-3}$, Kirkpatrick et al. 1994). An object like Kelu~1 
would appear to lie on the BD Pleiades sequence shown 
in Fig.~3 if it were at a distance of $\sim$90 pc. Since the 
the completeness limit of the 1 deg~$^2$ ITP survey was I$\sim$21, 
the probability to find one field object like Kelu~1 at a distance 
of 90$\pm$5 pc is in the range 25\%--10\%. Moreover, there were 
4 objects occupying the same region as Roque~25 in 
the $I,I-Z$ diagram of Zapatero Osorio et al. (1998). 
Hence, the probability 
that Roque~25 is a field object rather than a Pleiades member is less 
than 6\% .  We argue below that our spectroscopic observations support 
indeed that Roque~25 belongs to the cluster.    

For Roque~12, we obtain a heliocentric radial velocity 
of +9$\pm 12$~km s$^{-1}$ in the reference frame of the star VB10 
(v$_{rad}$=+35~km s$^{-1}$), consistent with cluster membership 
(Mart\'\i n et al. 1998; Stauffer et al. 1998).  
Our spectra of Roque~5 and 25 do not not allow us to obtain 
precise radial velocities because of their low S/N ratio. 
Another membership criterion is to consider gravity sensitive spectral 
indicators. Pleiades BDs should be typically younger 
than field VLM stars and BDs, and consequently they should have 
larger radii and lower gravities. Mart\'\i n et al. (1996) found 
that Teide~1 and Calar~3 have slightly weaker Na\,{\sc i} doublet 
(818.3~nm, 819.5~nm) and stronger VO bands than field 
stars of similar spectral type. Luhman et al. (\cite{luhman98}) 
reported that the Na\,{\sc i} lines are much weaker in an extremely 
young M8.5 BD belonging to the $\rho$Oph star-forming association than 
in field M8--M9 stars. 
In L-type objects the Na\,{\sc i} IR doublet becomes very weak 
because it is not a resonance line. In Roque~25, the Na\,{\sc i} lines are  
not detected. The absorption feature at $\sim$820~nm 
is probably due to a telluric H$_2$O 
band (Figure~1). On the other hand, L-type objects develop 
extremely broad  K\,{\sc i} resonance doublets (766.5~nm, 766.9~nm; 
Mart\'\i n et al. 1997; Tinney et al. 1998). Spectral synthesis 
calculations (Pavlenko 1997; Haushchildt 1998) show that the 
breath and strength of the K\,{\sc i} doublet is  
very sensitive to gravity and temperature. The K\,{\sc i} lines become 
narrower and weaker for lower gravity and/or higher temperature. 
We have found that the K\,{\sc i} 
doublet in Roque~25 is indeed weaker and narrower than in the field VLM star   
BRI~0021-0214. 
The PEWs measured with respect to the pseudocontinuum at 
$\sim$780~nm are given in Table~1. Due the large breath of the 
two K\,{\sc i} lines, their wings are blended and hence we measured  
the PEW of the whole doublet. 
The spectrum of Roque~25 is quite similar to that of the BD Kelu~1 
(Ruiz et al. 1997), as might be expected from their   
similar $I-K$ colors (Table~1). However, 
some differences are clear (Figure~1). 
Roque~25 has weaker  K\,{\sc i} doublet and 
stronger VO bands. We conclude that 
Roque~25 must have lower gravity than 
BRI~0021-0214 and Kelu~1 and hence it is  a younger, lower mass object. 
Thus, the gravity sensitive features in the spectrum of 
Roque~25 strongly support its membership to the young Pleiades cluster.   

The optical spectrum of Roque~25 is characterized by the following 
features: i) the lack of strong molecular absorption bandheads in the range 
640~nm--760~nm, in particular, the strong TiO bands starting at 705.0~nm 
that are conspicuous in Roque~12 and can still be seen 
in Roque~5, are absent or extremely weak in Roque~25; ii)  
the molecular systems of CaH, CrH and FeH become as strong or stronger 
than the systems of TiO and VO; 
iii) the atomic lines of K\,{\sc i} and Cs\,{\sc i} 
that are very strong in L-type objects (Ruiz et al. 1997; 
Mart\'\i n et al. 1997; 
Tinney et al. 1997, 1998) are weaker in Roque~25. 
Using the models of Baraffe et al. (1998) 
and the photometry of Zapatero Osorio et al. (1998),  
we infer T$_{\rm eff}\sim$2350~K and 2050~K for Roque~5 and Roque~25,  
respectively.    
The transition from 
M to L has to take place at a temperature slightly below that of Roque~5. 
Pleiades BDs with T$_{\rm eff}$ intermediate between Roque~5 and 25 should 
exist (the ITP, NOT and CFHT surveys have revealed a few candidates), 
and we expect that they will have a spectral type around 
L0 and T$_{\rm eff}\sim$2200~K. 
The temperature of Roque~25 estimated from the evolutionary models 
is warmer than the T$_{\rm eff}$ of 1900~K obtained 
by Ruiz et al. (1997) for Kelu~1, consistent with the bluer 
I-K color of Roque~25 (Table~1). Consequently, Roque~25 should  
be assigned a slightly earlier L-type subclass than Kelu~1. 
The calibration of the L-type temperature scale will have to await 
systematic observations 
of a large number of L-type objects and a good understanding 
of their atmospheres. We emphasize that the Pleiades BDs are   
very useful for calibrating the L-type class because the differences 
along the cluster spectral sequence should depend on mass and T$_{\rm eff}$, 
with age and metallicity being approximately constant (this is by no means 
the case for field objects). 

The chromospheric activity diminishes with decreasing mass along the 
Pleiades sequence. Roque~5 and 25  
do not show strong H$\alpha$ emission, consistent 
with the trend of decreasing chromospheric activity for decreasing 
temperature observed among Pleiades VLM members (Zapatero Osorio et al. 
1997). The same tendency is also present among field very late-M dwarfs 
(Basri \& Marcy 1995).  L-type field dwarfs do not usually have strong 
H$\alpha$ emission (Mart\'\i n et al. 1997; Tinney et al. 1998), although 
some exceptions do exist (Liebert et al. 1998). The lack of strong 
H$\alpha$ emission in Roque~25 indicates that L-type objects are not 
very active even when they are quite young.

In this work the Pleiades spectral sequence has been extended from  
the coolest M-types (Roque~4, Roque~5, PIZ~1), 
to the beginning of the L-type class (Roque~25). We have found 
that Roque~25 has photometric and spectroscopic properties supporting 
that it is a Pleiades L-type brown dwarf. 
The discovery of Roque~25 indicates that the cluster IMF  
extends to masses as low as $\sim$0.035~M$_\odot$.   
Very deep surveys should address the problem of whether there 
is a lower mass cutoff of the IMF. The minimum 
mass for BD-formation could be close to the 
deuterium burning mass limit ($\sim$0.012~M$_\odot$; cf. Burrows 
et al. 1997), as suggested by estimates of the minimum fragmentation scale 
of molecular clouds.  
The large $I-K$ color of Roque~25 indicates that near-IR large format 
cameras or mosaics would be advantageous to extend the Pleiades 
BD sequence to extremely low-mass objects. Future surveys should 
be sensitive to late L-type objects like those found in the field 
(Delfosse et al. 1997; Liebert et al. 1998; Kirkpatrick et al. 1998). 
L-type objects in the Pleiades would generally be less massive 
than their field counterparts because they are younger. 

\acknowledgments

{\it Acknowledgments}: 
This research is based on data collected at the 
W.~M. Keck Observatory, which is operated jointly by the University of 
California and Caltech, and on observations obtained at the William Herschel 
telescope, which is operated by the Isaac Newton Group funded by PPARC 
at the Spanish Observatorio del Roque de los Muchachos of the Instituto 
de Astrof\'\i sica de Canarias. We thank the staff of Keck observatory 
for carrying out service observations for us. We are grateful to I. Baraffe, 
P. Hauschildt and Ya. Pavlenko for sending theoretical computations prior 
to publication. 
We have made use of the Simbad database, operated at CDS, Strasbourg, 
France.   
This work was partially supported by the Spanish DGES under project PB95-1132-C02-01. 
EM acknowledges the support of  
the Spanish Ministry of Education and Culture. 
GB acknowledges the support of NSF through grant AST96-18439.

\clearpage

\figcaption[fig1bis.eps]{\label{fig1} 
Low resolution spectra of program objects. 
All of them have been normalized around 750~nm and are offset in steps 
of 2. A boxcar smoothing of 3 points has been applied to all the spectra. 
Note that the spectrum of Roque~5 has lower spectral resolution 
than the other three spectra (see text for details).  
Identification of the main spectral features is given in the top. 
The telluric absorption bands are marked with dashed lines.}

\figcaption[fig2.ps]{\label{fig2} The position of Roque~25 in the 
$I$ vs ($I-K$) sequence of Pleiades members. 
Open triangles represent the benchmark BDs 
Teide~1 (Rebolo et al. 1995, 1996) and Roque~4 (Zapatero Osorio et al. 1997). 
Open circles denote other Pleiades objects. 
The dashed line corresponds to a 120~Myr theoretical isochrone 
(Baraffe et al. 1998) and is labelled with masses in solar units. 
The solid line is the sequence of field VLM dwarfs (Leggett, Allard \& 
Hauschildt 1998) 
shifted to the Pleiades distance (m-M=5.53). The open squares are the 
field stars BRI~0021-0214 and GD165B also shifted to Pleiades distance.}

\clearpage

\begin{deluxetable}{lccccc}
\footnotesize
\tablecaption{\label{tab1} Data for program objects}
\tablewidth{0pt}
\tablehead{
\colhead{Parameter}  & 
\colhead{Roque 12} &
\colhead{Roque 5} &
\colhead{Roque 25} &
\colhead{BRI 0021-0214} &
\colhead{Kelu 1} 
                              }
\startdata
$I_{C}$ & 18.47 & 19.71 & 21.17 & 15.1  & 16.8   \nl
$I_{C}-K_{UK}$ & 3.37 & 4.31 & 4.90 & 4.43 & 4.98  \nl
SpT & M7.5 V & M9 V & early L & M9.5 V & early L \nl
PEW (H$\alpha$) (\AA) & 19.7$\pm$0.3 & $\le$8 & $\le$5 & $\le$1 & 0.9$\pm$0.6 \nl
PEW (Li\,{\sc i}) (\AA) & $\le$1.5 & $\le$8 & $\le$5 & $\le$0.4 &  
4.3$\pm$0.6 \nl
PEW (K\,{\sc i}) (\AA) & 26$\pm$3 & 35$\pm$10 & 
48$\pm$5 & 63$\pm$6 & 94$\pm$8 \nl
\enddata
\tablenotetext{}{The PEW  of H$\alpha$ refer to line emission with 
respect to the local pseudocontinuum, whereas 
PEW (Li\,{\sc i}) and PEW (K\,{\sc i}) refer to line absorption. 
}
\end{deluxetable}

\clearpage


\plotone{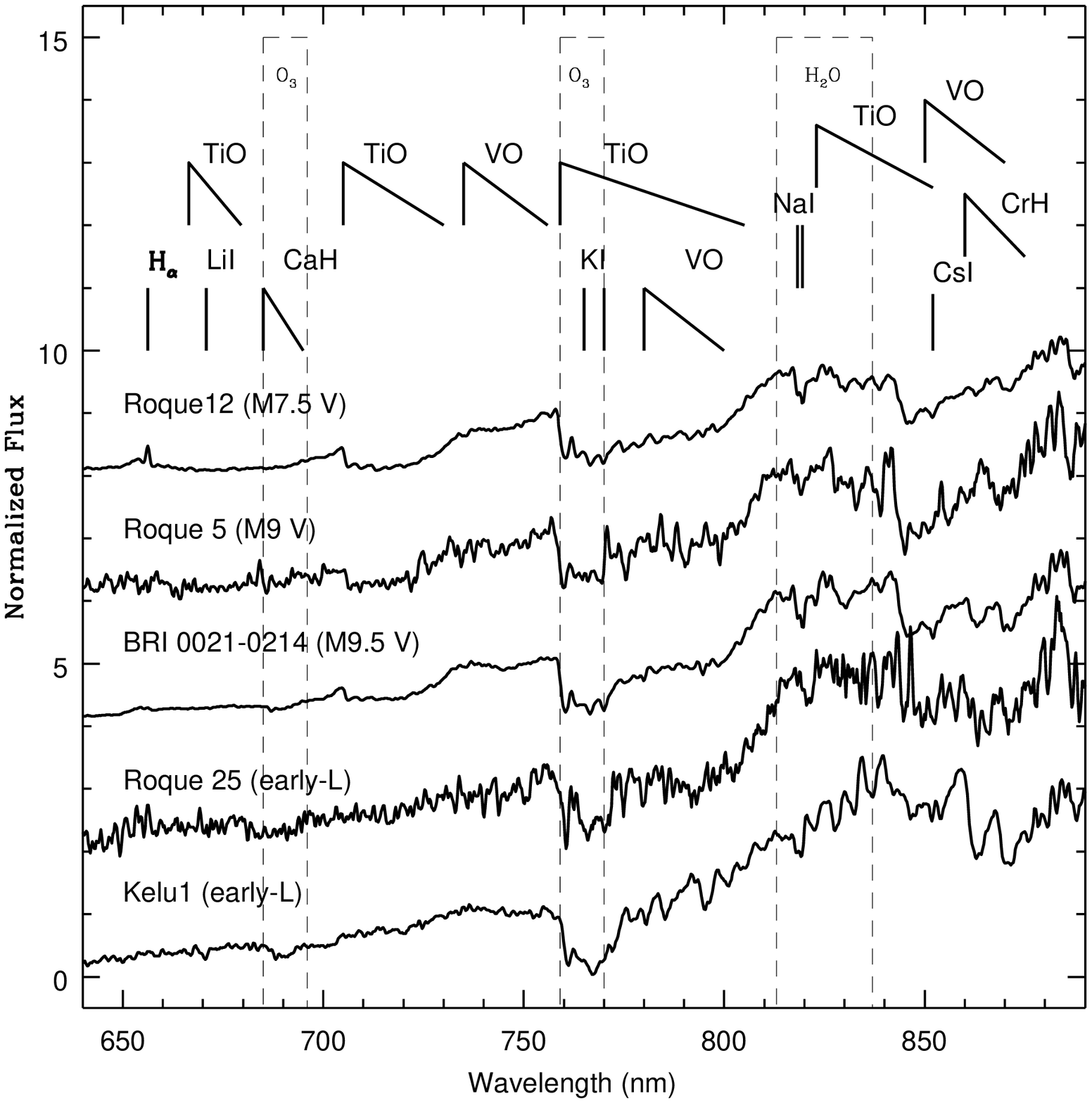}

\plotone{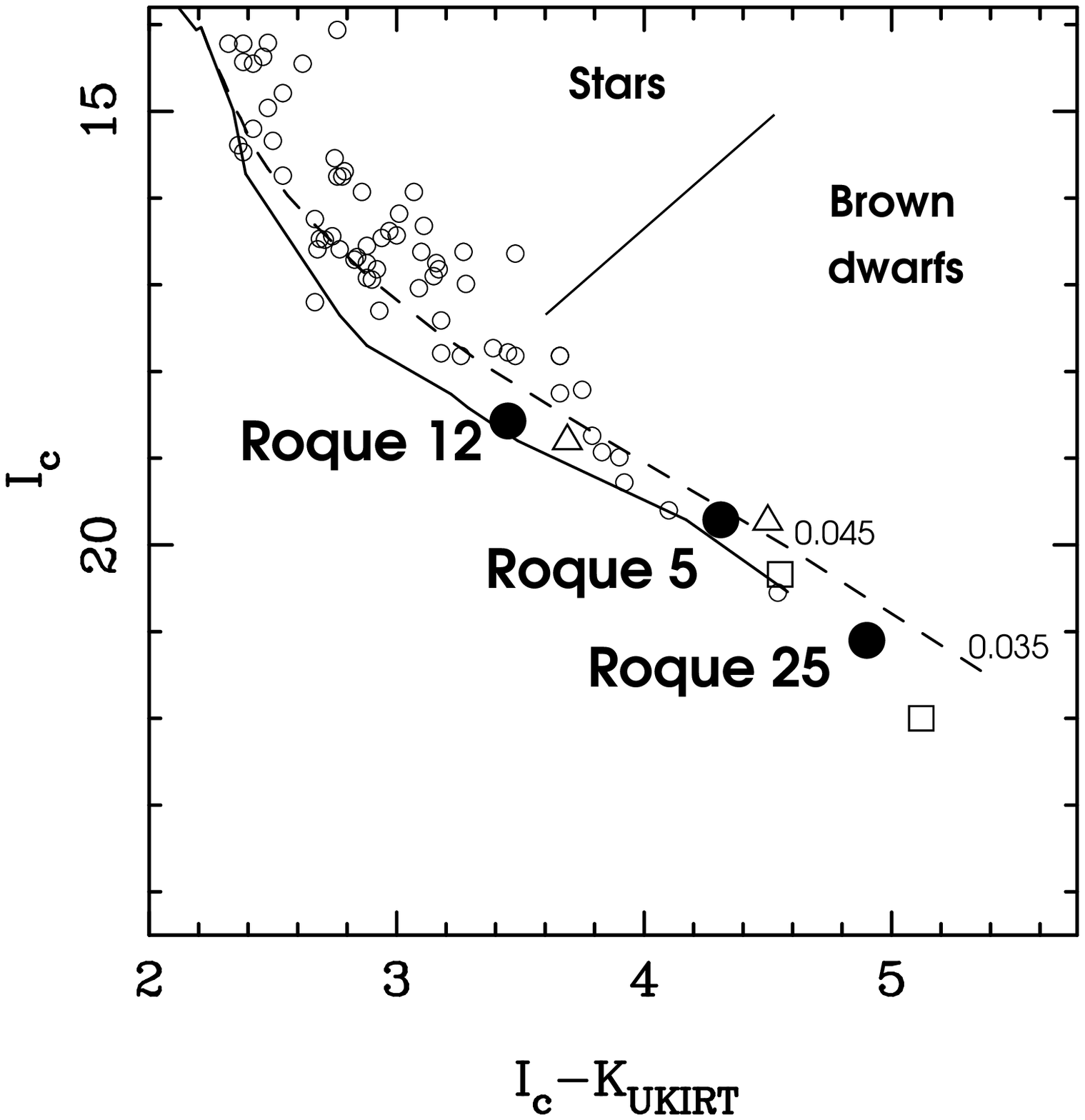}

\end{document}